\documentclass[journal,twoside]{IEEEtran}
\usepackage{graphicx}

\ifCLASSINFOpdf
\else
\fi
\begin{document}
%
\title{Profile and Crowding of Currents in Mesoscopic Superconductors with an Array of Antidots}
%
%
%

\author{D.\ Okimoto, E.\ Sardella, R.\ Zadorosny
\thanks{D.\ Okimoto and R.\ Zadorosny are with Grupo de Desenvolvimento e Aplica\c{c}\~{o}es de Materiais, Faculdade de Engenharia, \textit{UNESP--Universidade Estadual Paulista}, Departamento de F\'{\i}sica e Qu\'{\i}mica, Caixa Postal 31, 15385-000, Ilha Solteira, SP, Brazil.}
\thanks{E.\ Sardella is with {Faculdade de Ci\^{e}ncias,  \textit{UNESP--Universidade Estadual Paulista}, Departamento de F\'{i}sica, Caixa Postal 473, 17033-360, Bauru, SP, Brazil.}}}

%
%

{OKIMOTO \MakeLowercase{\textit{et al.}}: Profile and Crowding of Currents in Mesoscopic Superconductors with an Array of Antidots}
%



\maketitle

\begin{abstract}
Studies with mesoscopic superconducting materials have made significant advances on the last decades. One of the applications of such systems is in devices for single photon and single electron detectors. However, depending on the geometry of these systems, crowding current effects take place and, as a consequence, the total critical current could decrease which facilitates the penetration of vortices. This effect could also be responsible for a variety of penetration morphologies of flux avalanches in macroscopic samples. Thus, in this work we used the time-dependent Ginzburg-Landau theory to study the crowding current effects in mesoscopic superconducting systems with an array of antidots. It is demonstrated that the profile of the currents is influenced by the antidots, i.e., in the vertices of the antidots the intensity of the currents increases. On the other hand, the profile of the currents in between the antidots is more affected in the smaller system which presents closer antidots.
\end{abstract}

\begin{IEEEkeywords}
TDGL, mesoscopic, crowding current, antidots.
\end{IEEEkeywords}

%
\IEEEpeerreviewmaketitle

\section{Introduction}
%
%
%
%



\label{Intro}

The advances of nanofabrication techniques on the last decades stimulated significantly studies with materials in the meso and nanometric scales. Particularly, superconducting materials with such sizes are known as mesoscopic systems and exhibit a variety of behaviors which arise due to confinement effects \cite{connoly,schweigert,melnikov,berdiyorov,muller,sardella4,palacios,misko,zhao2,cren,golubovic}. As a consequence, such specimens can be used in applications like single photon \cite{goltsman,kerman,dorenbus} and single electron \cite{rosticher} detectors, amplifiers \cite{eom} and imaging of single magnetic quantum flux (single vortex) \cite{vasyukov}.

Another possible application is related to the control of the vortex penetration and motion by such mesoscopic systems since this can avoid dissipation and, consequently, should increase the critical current density, $J_c$, and the upper critical field, $H_{c2}$ of the material. However, an effect described by Hagedorn and Hall in 1963 \cite{hagedorn}, i.e., the crowding current (CC) effect, initially applied to non-superconducting materials, can interfere on the functioning of the cited devices.

The CC effect is characterized by an agglomeration of the current lines in the regions where such current needs to suddenly change its flow direction, as to contour edges and defects. This crowding of the current can cause a decreasing of the total critical current of the system ~\cite{jones,clem1,clem2,henrich,vestgarden,hortensius,adami} and, as a consequence, these regions became good channels for the vortex penetration and the beginning of a resistive state.

Some authors have already reported that the CC effect is also responsible for a variety of morphologies acquired by dendritic penetrations in systems with antidots, ADs, with different geometries \cite{clem1,clem2,adami,zadorosny,motta}. In such case, the crowding of the current maximizes the Lorentz force over the magnetic flux trapped in the ADs and then guides the propagation of the flux avalanches by depinning the vortices.

In the present work we used the time-dependent Ginzburg-Landau theory to analyze the vortex dynamics and the current crowding effect in mesoscopic superconducting materials with a square array of sixteen square antidots of fixed size. We varied the distances between the antidots and, consequently, the size of the systems. Thus, we analyzed the crowding of the currents in the vicinity of the ADs.

\section{Theoretical Formalism}
\label{TDGL}

In the Ginzburg-Landau, GL, formalism the superconducting state is described locally by a complex order parameter $\psi$ which the value of $|\psi|^2$ represents the density of Cooper pairs. In the case of type II superconductors the nucleation of vortices inside the material becames energetically favorable. In the presence of an external field, flux could penetrates the material which causes the appearance of superconducting currents to shield the material. Then, to determine the local field, ${\bf h}$, the order parameter, $\psi$, and the superconducting current density, ${\bf J}_s$, in the system, the normalized time dependent Ginzburg-Landau equations, TDGL, were used:

\begin{eqnarray}
\left ( \frac{\partial}{\partial t} + i\varphi\right ) \psi& = & -\left
(-i\mbox{\boldmath
$\nabla$}-{\bf A} \right )^2\psi\nonumber \\
& & +(1-T)\psi(1-|\psi|^2)\;,\nonumber \\
\beta\left ( \frac{\partial{\bf A}}{\partial t}+\mbox{\boldmath
$\nabla$}\varphi \right ) & = & {\bf J}_s-\kappa^2\mbox{\boldmath
$\nabla$}\times{\bf h}\;,\label{tdgleq}
\end{eqnarray}
where the supercurrent density is given by
\begin{equation}
{\bf J}_s=(1-T){\rm Re}\left [ \psi^{*}\left ( -i\mbox{\boldmath
$\nabla$}-{\bf A} \right )\psi \right ]\;, \label{densityCurrent}
\end{equation}
${\bf A}$ is the vector
potential which is related to the local magnetic field as
${\bf h}=\mbox{\boldmath $\nabla$}\times{\bf A}$, and $\varphi$ is the scalar potential. With the normalization, all the distances are in units of the coherence length
at zero temperature $\xi(0)$; the magnetic field is in units of the zero temperature upper critical field
$H_{c2}(0)$; the temperature $T$ is in units of the critical
temperature $T_c$; the current is in units of of the depairing current $J_0$; the time is in units of the characteristic
time $t_0=\pi\hbar/8k_BT_c$; $\kappa$ is the Ginzburg-Landau parameter;
$\beta$ is the relaxation time of $\bf A$, related
to the electrical conductivity.

The TDGL equations  are gauge invariant under the transformations $\psi^{\prime}=\psi e^{i\chi}$,
${\bf A}^{\prime}={\bf A}+\mbox{\boldmath $\nabla$}\chi$,
$\varphi^{\prime}=\varphi-\partial\chi/\partial t$ \cite{gropp}. Thus, we conveniently choose the zero-scalar potential gauge, that is,
$\varphi^\prime=0$ at all times and positions.

\section{Results and Discussion}
\label{results}

\begin{figure}
\centering
\includegraphics[width=0.7\columnwidth,height=1.2\linewidth]{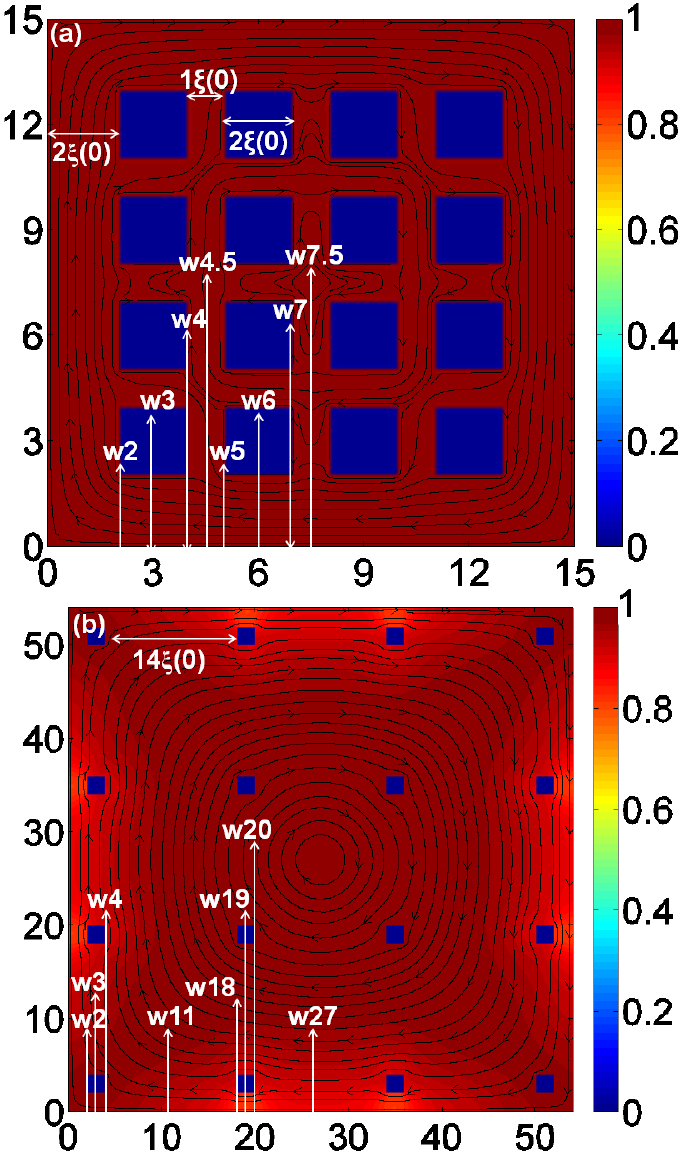}
\caption{(Color online) Simulated systems (a) $S_1$ and (b) $S_2$. The vertical arrow and the notation "$w\*$" indicate the distance counted from the left border of the system where the modulus of ${\bf J}_s$ was analyzed. The ADs were inserted at a distance of $2\xi(0)$ from the borders.  However, the distance between them was changed by increasing the size of the system such that $d= 1\xi(0)$ for $S_1$ and $d= 14\xi(0)$ for $S_2$.}
\label{fig1}
\end{figure}

In our simulations, we worked with $\kappa = 5$, which is equivalent to a Pb-In alloy \cite{poole}. Thus, to study the effects of an array of ADs in mesoscopic superconductors, two square systems with different lateral sizes, $L= 15\xi(0)$ (which will be labeled $S_1$) and $54\xi(0)$ (which will be labeled $S_2$), were studied. In those systems, square ADs with lateral size of $2\xi(0)$ were disposed in a square array which counted with sixteen specimens, as shown in Fig.~\ref{fig1}. The distance between the border of the systems and the ADs was maintained fixed as $2\xi(0)$. On the other hand, the distances $d$ between the ADs were changed by increasing  $L$. Fig.~\ref{fig1} shows the two simulated systems. The vertical arrows indicate the distance from the left border of the system where the modulus of $J$ was analyzed, i.e., the notation $w\#$ indicates the distance (in units of $\xi(0)$) from the left border of the system along the $x$ axis. Then, in Fig.~\ref{fig1}(a) is shown the $S_1$ system and in Fig.~\ref{fig1}(b) is shown the $S_2$ system. The black lines indicate the circulation of ${\bf J}_s$. For comparison, all the analysis were carried out in the Meissner state and at a fixed temperature and external applied magnetic field, $T=0.8T_c$ and $H=0.23H_{c2}(0)$, respectively.

\begin{figure}
\centering
\includegraphics[width=0.7\columnwidth,height=1.2\linewidth]{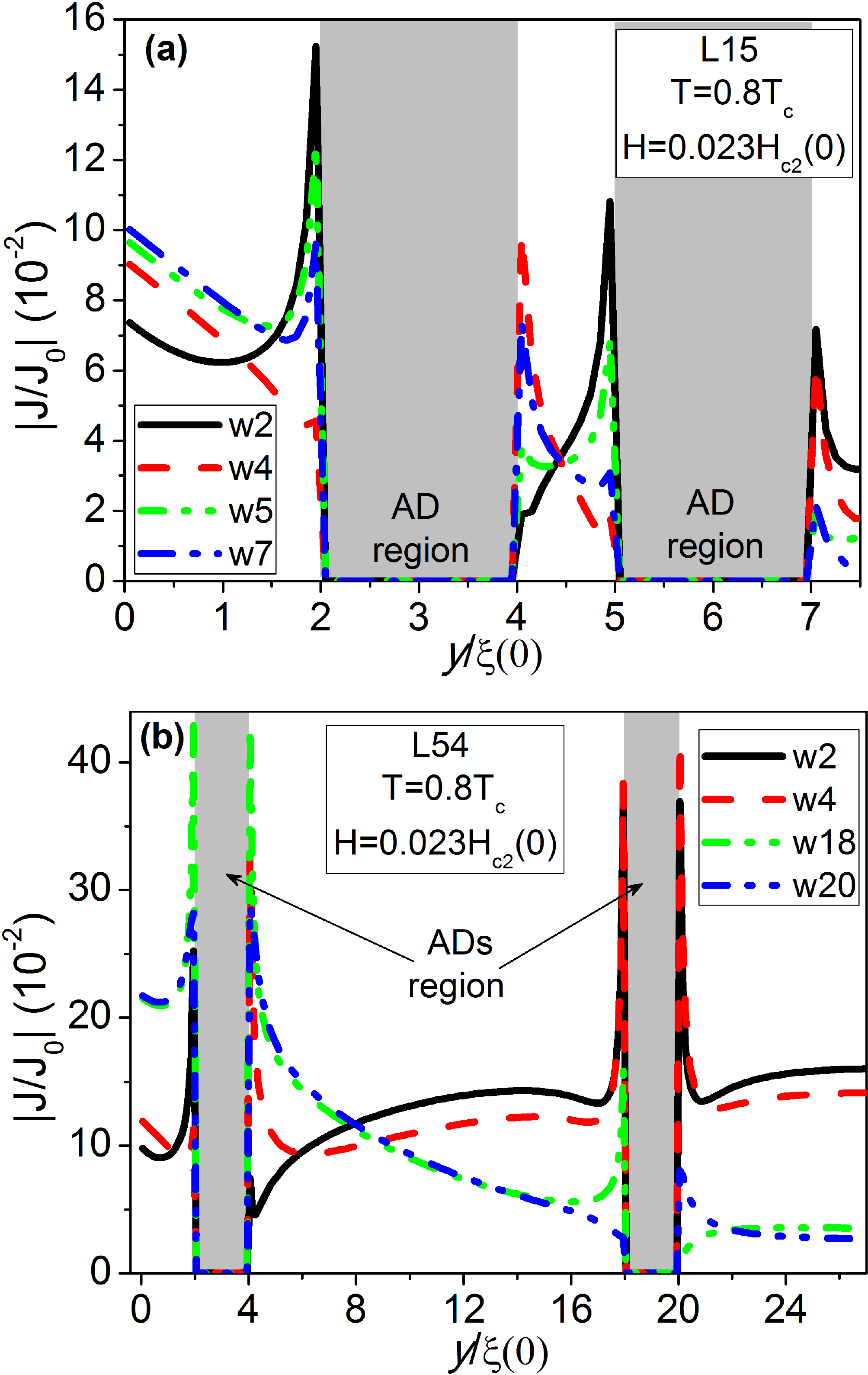}
\caption{(Color online) Modulus of ${\bf J}_s$ along the $y$ axis and for fixed positions for $x$ coordinate as illustrated in Fig.~\ref{fig1}. The currents are amplified in the most of the ADs vertices due to crowding effects. The smaller system $S_1$ (panel (a)) is more affected by the decreasing of $J_s$ than the larger one $S_2$ (panel (b)) as can be seen by the evolution of the peaks in the vicinity of the vertices of different ADs.}
\label{fig2}
\end{figure}

In order to analyze how the profile of the modulus of ${\bf J}_s$ is be affected by the ADs, we plotted $J(y)$ at fixed positions $w\#$. Fig.~\ref{fig2} shows such curves which were taken in the vicinity of the border of the ADs. Following the square symmetry of the system, the curves were ordered from outer to the inner border of the ADs. In general, and in both $S_1$ and $S_2$ systems, the currents are amplified at the vertices of the ADs as expected due to the CC. However, such amplification does not occur in all vertices which could be associated with the distance between the ADs and with the fact that the distribution of $J_s$ decreases in the deeper regions in the system. The $J_s$ distribution is also distinguishably affected in the superconducting region in between the ADs for $S_1$ and $S_2$. This behavior could be related to the different distances of separation of the ADs, as can be verified in panels (a) and (b) of Fig.~\ref{fig2}. An interesting behavior is shown in panel \ref{fig2}(b) where $J_s$ increases in between the ADs near the border and decreases in between the inner ADs. Despite this behavior, the amplitude of $J_s$ is quite the same in the vertices of the outer and inner ADs.

\begin{figure}
\centering
\includegraphics[width=0.7\columnwidth,height=1.2\linewidth]{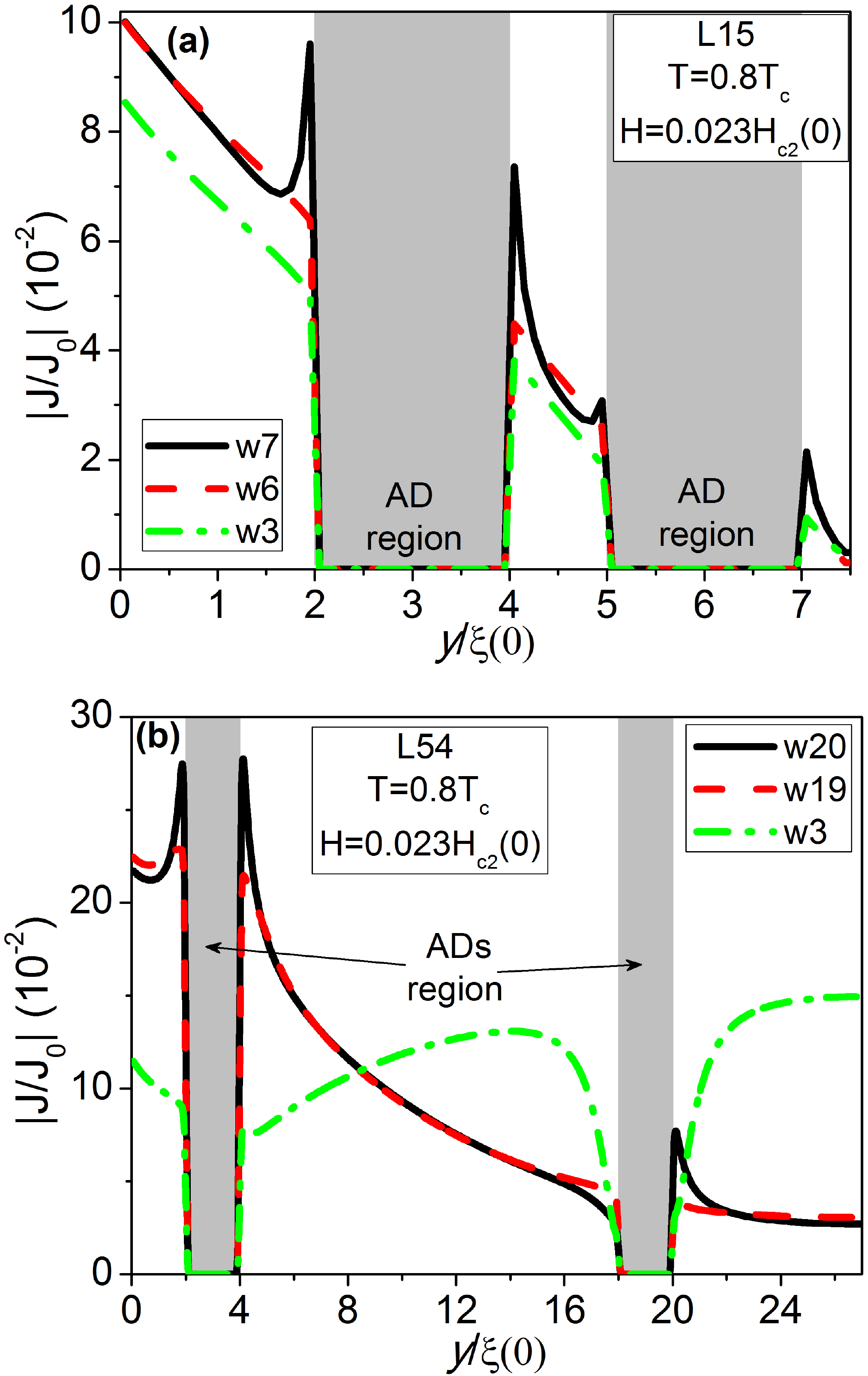}
\caption{(Color online) Modulus of $J_s$ along the $y$ axis at fixed position of the $x$ coordinate which passes through the center of the ADs. In comparison with the curves which pass in the vicinity of the ADs border - $w7$ for $S_1$ and $w20$ for $S_2$ - in such regions there is no presence of the crowding effects.}
\label{fig3}
\end{figure}

In Fig.~\ref{fig3} is shown the profile of $J_s$ along the line which crosses the ADs in the middle, as denoted by the curves $w3$ and $w5$ for $S_1$ and $w3$ and $w19$ for $S_2$. The curves $w7$ and $w20$ are inserted only for comparison purpose  with the crowding effects which occur in the vertices. In general, there is no such effect in those regions. On the other hand, the presence of ADs do not affect the general behavior of $J_s$ in those positions for both systems. The exception is the curve $w3$ for $S_2$ (panel \ref{fig3}(b)) for which $J_s$ increases in between the ADs. This behavior should be related to the proximity to the border of the system where $J$ increases as approximates to the middle of the lateral side ($y=L/2$).

\begin{figure}
\centering
\includegraphics[width=1\columnwidth,height=0.7\linewidth]{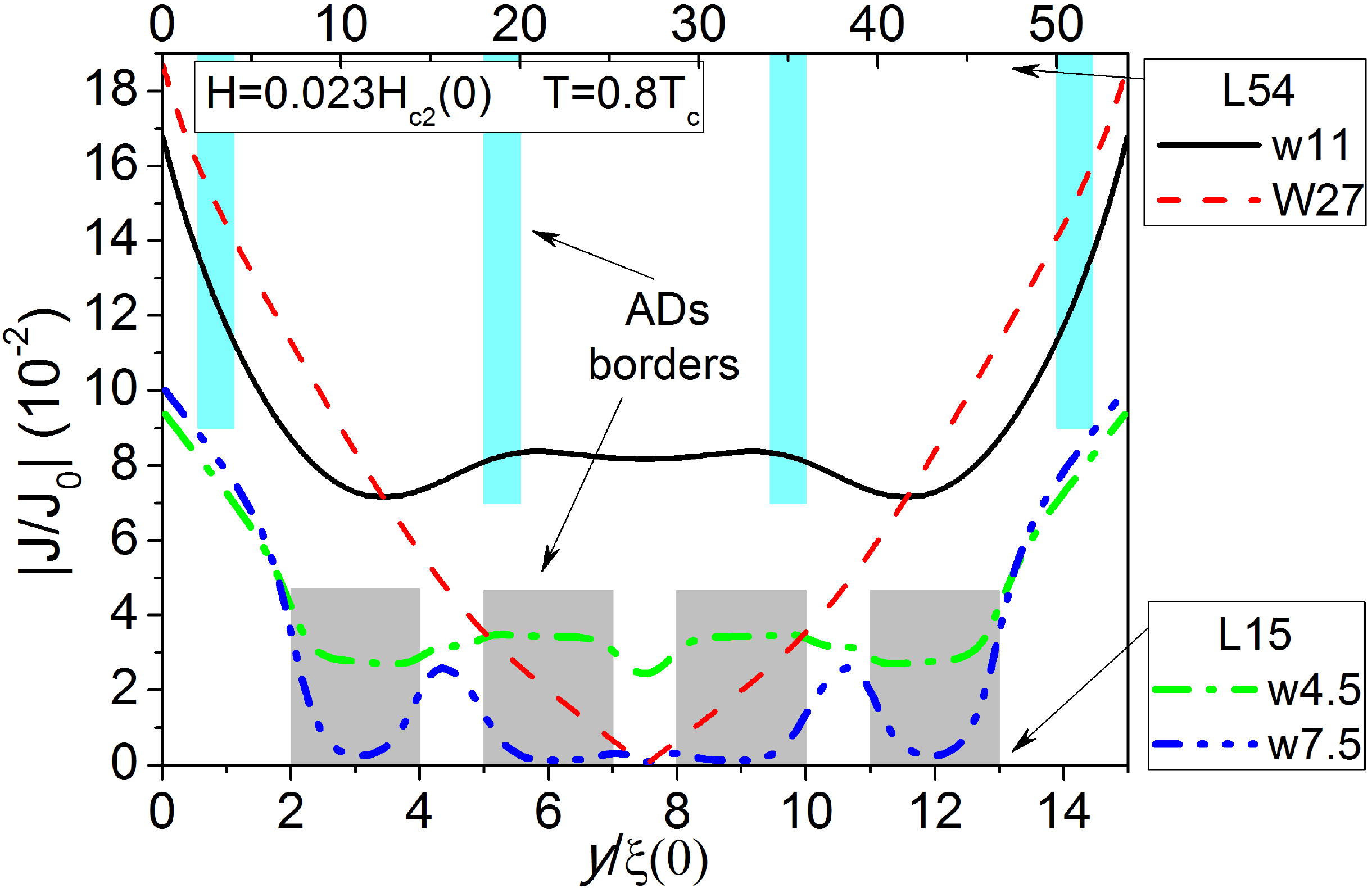}
\caption{(Color online) Modulus of $J_s$ along the $y$ axis at fixed position of the $x$ coordinate which passes in between the ADs. The profile of the current is influenced by the ADs in $S_1$, but not in $S_2$. Such behavior is due to the proximity of the ADs, as indicated by the bars.}
\label{fig4}
\end{figure}

The profile of $J_s$ in the region between the ADs are shown in Fig.~\ref{fig4}. The bars in such figure indicates the region which has an AD in the neighborhood. The curves of $w4.5$ and $w11$ for $S_1$ and $S_2$ respectively, are those in the vicinity of the left border of the system. Both curves present a similar behavior for which $J_s$ does not vanish at the position $y=L/2$. Nonetheless, the curves $w7.5$ and $w27$ vanish at the center of the system. Besides some similar aspects, the profile of $J_s$ is clearly influenced by the proximity of the ADs in $S_1$, but not by those of $S_2$.

\section{Conclusion}

In this work we analyzed the profile of the modulus of the current in two mesoscopic superconducting systems with an array of ADs. It was shown that the crowding of the currents occurs in the vertices of the ADs even with the decreasing of $J$ as the center of the systems are reached. The profile of $J$ in the smaller system is influenced by the ADs stronger than in the bigger system. In general, the crowding effect in the vertices of the ADs should be the responsible for the guidance of the flux penetration in vortex avalanches. Such systems should be used in future fluxonics devices. The behaviors demonstrated in this work can also be extended to real thin films since the currents are distributed along the entire system.


%



\section*{Acknowledgment}


We thank the Brazilian Agencies Fundunesp/PROPe grant 2115/002/14-PROPe/CDC and the S\~{a}o Paulo Research Foundation (FAPESP), grant 2013/17719-8 for financial support.

\ifCLASSOPTIONcaptionsoff
  \newpage
\fi

\end{document}